\documentclass[twoside]{article}
\usepackage[latin1]{inputenc}
\usepackage{t1enc}
\usepackage{a4}
\usepackage{tabularx}
\usepackage{epsf}

\def\bref{\vspace{4pt}\noindent\hangindent=10mm}

\def\arcmin{\hbox{$^\prime$}}
\def\arcsec{\hbox{$^{\prime\prime}$}}

\def\sun{\hbox{$\odot$}}
\def\kms{\hbox{km\,s$^{-1}$}}

\def\Msun{\hbox{M$_{\sun}$}}
\def\Ha{\hbox{H$_{\alpha}$}}

\begin{document}
\setcounter{figure}{0}
\setcounter{section}{0}
\setcounter{equation}{0}

\begin{center}
{\Large\bf
Warm and Hot Diffuse Gas in Dwarf Galaxies
}\\[0.7cm]
Dominik J. Bomans \\[0.17cm]
Astronomy Institute, Ruhr-University Bochum \\
Universit\"atsstr. 150, 44780 Bochum, Germany\\
bomans@astro.ruhr-uni-bochum.de http://www.astro.uni-bochum.de/bomans
\end{center}

\vspace{0.5cm}

\begin{abstract}
\noindent{\it
Dwarf galaxies provide a special environment due to their low mass, small 
size and generally low metal content. These attributes make them perfect 
laboratories for the interaction of massive stars with the interstellar 
medium on small and especially large spatial scales.  The natural result 
of the spatially concentrated energy output from stellar winds and supernovae 
of an OB association is an expanding bubble. These bubbles can grow to 
kpc-size and become the 
dominant driver of the chemical and dynamical evolution of dwarf galaxies. 
In such low mass systems, bubbles have an enhanced probability of breaking 
out of the gaseous disk into the halo of the host galaxy. This may lead to 
venting metal enriched hot gas to large distances from the sites of creation. 
In this work I review the current observational material on hot gas inside 
bubbles, blow-outs, and hot halos of dwarf galaxies and discuss several 
conclusions which can be drawn from the observations. I will also present an 
analysis of the dwarf galaxy NGC\ 1705 as a case study, highlighting 
observational methods and problems with the current data. Finally I will 
comment on some areas where large progress should be possible in the near 
future. 
}
\end{abstract}

\section{Introduction}
Dwarf galaxies, by definition of their class, are galaxies of low 
mass and size. This directly implies that they have a much 
weaker gravitational potential well than typical spiral galaxies and 
fill a smaller volume with gas and stars.   All dwarf galaxies, even very 
low mass systems, show quite complicated star formation histories (e.g. 
Mateo 1998). From the color-magnitude diagrams one derives the presence 
of some low level 
star formation activity over most of the lifetime, sometimes interrupted by 
short intervals of strongly enhanced star formation rate. These results 
and the presence of blue 
compact dwarf galaxies, where the current star formation rate 
is so high that these dwarf galaxies appear as isolated giant HII regions, 
tells us that bursts of star formation are indeed a natural process in dwarf 
galaxies, as predicted by models of stochastic self 
propagating star formation (Gerola et al.\ 1980).  Bursts of star 
formation, meaning spatially and temporally correlated energy 
input from massive stars and supernova explosions inside physically small 
systems, lead to a strong response of the gas in the host galaxy. 


Analytical and numerical 
modeling of the reaction of the gas to the energy input from stellar winds 
and supernovae is an active topic since the papers of Castor et al. (1975) and 
Weaver et al. (1977). Recent examples are e.g. Freyer \& Hensler (2000), 
Strickland \& Stevens (2000), Mac Low \& Ferrara (1999), and Tomisaka (1998).
For a review of the basic ideas, see Tenorio-Tagle \& Bodenheimer (1988).
The result of the energy input into the interstellar medium (ISM) is basically 
an expanding bubble of hot gas inside the substrate of cool gas of 
the host galaxy. The hot bubble is enclosed by a dense cool shell, which has
ionized gas at the inner boundary layer between the hot gas and the shell.  
If the bubble grows to a linear diameter comparable to the neutral 
gas scale height of the galaxy, the expansion speeds along the z-axis 
(e.g. Mac Low \& McCray 1988) and the shell expands into the lower halo of 
the host galaxy while starting to deform and break due to Rayleigh-Taylor 
instabilities (e.g. Mac Low et al. 1989). 

Since massive stars and supernova explosions are the dominant source of 
heavy elements, the newly processed material is located inside the 
shells.  Whether and to which degree it is moved upward out of a 
galaxy and what happens to this gas in the lower halo, is of crucial 
importance for the understanding of the chemical evolution of 
dwarf galaxies (e.g. Hensler \& Rieschick 1999). In the work Mac Low \& 
Ferrara (1999) for the first time a set of hydrodynamical simulations 
incorporating a relatively detailed model of the 
dwarf galaxy potential (including dark matter) was used and the 
simulations for a whole dwarf galaxy was run for more that 100 Myr.  
Still, the authors had to compromise e.g. by relatively 
basic treatment of the cooling processes and ignoring magnetic fields.

The hot gas inside the bubbles is predicted to be in the temperature 
range of $10^5$ and $10^7$ K (e.g. Weaver et al. 1977), which implies that 
the plasma will radiate in the extreme UV and soft X-ray regime.  The ionized 
gas at the boundary layer between the hot interior and the cool shell wall 
should be visible in optical and UV emission lines, while the cool shell
itself can be observed in 21cm emission. 
Since density of the substrate medium and especially the size of the energy 
depositing stellar association (from few O or even B stars to many thousands 
of OB stars as e.g. inside giant HII regions like 30 Dor or NGC 5471 in M101)
span a large parameter space, the size of the bubbles can vary a lot, 
from pc to kpc scale (Chu 1995).   
The (more abundant) small bubbles do not break out of the disk of a galaxy. 
These bubbles do still structure the interstellar medium and should lead to 
ionized filaments, as observed 
in the Magellanic Clouds (e.g. Kennicutt et al. 1995) and the Milky Way 
(Haffner et al.\ 1999). They also lead to large regions of hot gas observed 
in the LMC (e.g. Chu \& Mac Low 1990, Bomans et al. 1994) and the 
Milky Way (e.g. Snowden et al. 2000).

The observation of warm and hot gas in dwarf galaxies allows therefore 
to study the mechanism shaping the topology and phase structure of the 
ISM as well as the processes responsible for the chemical and (at least 
partly) the dynamical evolution of the host galaxy.  Dwarf galaxies are 
supremely suited for this task.  They present the most extreme environment 
for feedback of massive stars on the interstellar medium 
due to their shallow potential wells, their small sizes, and the absence 
of complicating other factors like density waves.

\section{OB associations and the interstellar medium}

\begin{figure}[t]
\epsfxsize=0.80\textwidth
\centerline{\epsffile{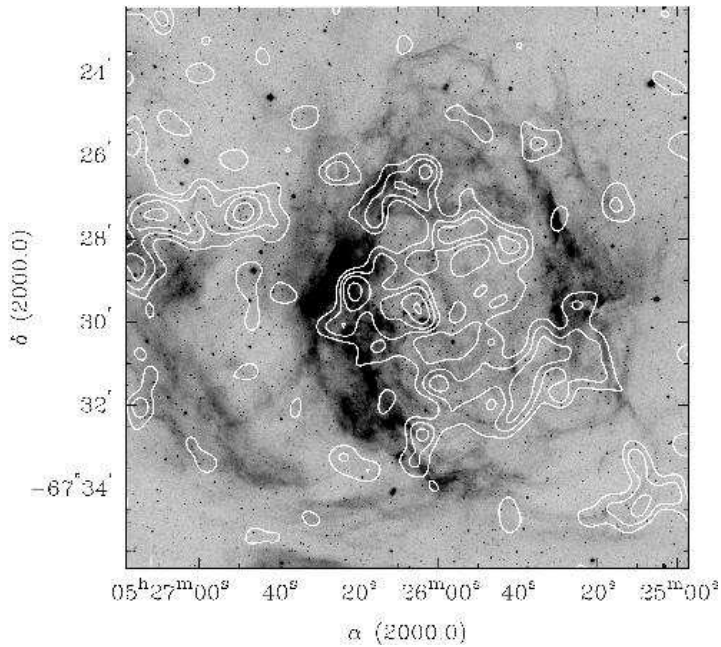}}
\caption{\Ha\ image of the LMC superbubble N51D with contours of the ROSAT 
soft and medium band X-ray emission overplotted.} 
\label{fig:n51d_x}
\end{figure}

Before going into details of the properties of diffuse X-ray emission of 
dwarf galaxies, it may be fruitful to check a few of the assumptions and 
theoretical predictions in smaller, better controlled, nearby systems, like 
the superbubbles in the Magellanic Clouds.  Fig.\,\ref{fig:n51d_x} shows an 
\Ha\ image of the superbubble N51D in the Large Magellanic Cloud (LMC). 
The structure is a fairly typical example of its class (Meaburn 1980). It 
shows a complete shell of ionized gas with a diameter of about 100 pc 
centered on one large OB association.
Such a superbubble distinguishes itself physically from a ($\sim 1 - 50$ pc) 
bubble around a single star (Weis \& Duschl 1999) and a kpc-sized 
supergiant shell (sometimes also called supershell) (Meaburn 1980; Chu 1995) 
around a large complex of OB associations.

N51D expands with a velocity of $\sim 35$ \kms (Lasker 1980; de Boer \& Nash 
1982) into the surrounding medium.  Despite of some larger loop-like 
extensions of the \Ha\ shell to the north and the south-west (as visible in 
Fig.\,\ref{fig:n51d_x}), this 
superbubble appears to be still a closed volume.  The difference in surface 
brightness between the eastern and western section is partly due a different 
density of the surrounding gas into which the superbubble expands.  The 
gas is ionized by the OB association inside, which contains one the  
UV brightest stars in the LMC, the WC5+O8Iaf binary Sk-67 104 (= HD 36402).  
A detailed study of the OB association (Oey \& Smedley 1998) yields an age of 
3 Myr and determines from the observed stellar content the total 
energy input of the stars to the superbubble.   N51D emits diffuse 
X-ray emission (Chu \& Mac Low 1990) which is brighter near the eastern 
rim of the superbubble.  Fig.\,\ref{fig:n51d_x} shows the \Ha\ image of N51D 
overlayed 
with contours of the X-ray emission derived from an ROSAT PSPC archival data
(Bomans et al. 2001a, in prep).
While some X-ray emission inside N51D seems to stem from point sources 
(most probably X-ray binaries), diffuse emission is clearly present inside 
the cavity delineated by the \Ha\ shell.  The ROSAT data confirm the 
surface brightness enhancement at the bright, probably denser, eastern 
boundary of N51D first detected using EINSTEIN (Chu \& Mac Low 1990), 
but also show considerable substructure here, not following one-to-one 
the \Ha\ emission (Bomans et al.\ 2001a, in prep.).  The temperature of the 
hot gas is about $5 \times 10^6$ K, when fitting spectra with collisional 
equilibrium models (e.g. Raymond \& Smith 1977).  The X-ray surface brightness 
and luminosity of N51D does not fit well to the predictions of the 
Weaver et al.\ (1977) model of an expanding superbubble and require an 
additional source of X-ray emission, which can be provided by supernova 
explosions of 
massive stars in the OB association, meaning inside the superbubble.  
As soon as the shock wave hits the dense shell wall, hot gas with high density 
and therefore surface brightness is produced, a process we may well observe 
right now at the eastern inner shell wall (Chu \& Mac Low 1990, Bomans et al.\
2001a, in prep.).
It is interesting to note here, that the loop at 
the south-western edge of N51D is a separate diffuse X-ray region, which 
is tempting to be identified with a beginning outflow, similar 
to the one in the LMC superbubble N44 (Chu et al.\ 1993). 
With the availability of an HI synthesis map of the LMC (Kim et al.\ 1998), 
even some informations about the surrounding neutral gas density and 
topology can be derived.   
Therefore in the case of N51D all critical parameters for the evolutions 
of a bubble can be observationally determined or at least estimated.  

The crucial determination of the metallicity of the diffuse hot gas is still 
missing.  The current information on superbubbles does not 
give reliable metallicities or how metal-enriched and more metal-poor gas is 
distributed in the bubble and along the shell walls, much less than giving 
details on the actual mixing processes.
ASCA observations of a sample of LMC supernova remnants prove at least the 
first assumption: the hot gas is indeed metal-enriched by the supernova 
explosion, but it shows also that mixing of hot metal-rich and cool metal-poor 
gas starts right away (Hughes et al. 1998).

\section{Hot Diffuse Gas}

\subsection{Observations}
The number of dwarf galaxies, which had their X-ray properties investigated 
is still relatively small.  Of all galaxies observed with the 
EINSTEIN satellite (Giacconi et al.\ 1979) and compiled by 
Fabbiano et al. (1992), 8 dwarf galaxies are present (see 
Tab.\,\ref{tab:x_dwarfs}).
This list does not include the obvious cases of the LMC (e.g. Chu \& Mac Low 
(1990), Wang et al.\ (1991)) and the SMC (Wang \& Wu 1992). These two galaxies 
have small enough distances to study diffuse sources like 
supernova remnants or superbubbles individually. Only point sources were 
detected in the other dwarf galaxies of the Fabbiano et al. (1992) atlas with 
no convincing case of extended X-ray emission.  

The sensitivity and spatial resolution improvement ROSAT 
(Tr\"umper 1993) provided over EINSTEIN changed the conditions significantly, 
but still the number of observed and analyzed dwarf galaxies is 
still quite low, as demonstrated in Tab.\,\ref{tab:x_dwarfs}.  Here the 
relevant X-ray data of all dwarf galaxies with analysis of their X-ray 
emission are compiled.  Due to its large luminosity, mass and clear disk 
structure, M\ 82 is not regarded as dwarfish galaxy and is therefore not 
included in this list.

\begin{table}
\caption[]{X-ray detected dwarf galaxies}
\label{tab:x_dwarfs}
\begin{center}
\begin{tabular}{ccccccc}
\hline
\hline
galaxy        & D     & M$_B$ & point & diffuse &  kT & references \\
              & [Mpc] &       &        &       & [keV] & \\
\hline
\hline
LMC           & 0.05& -17.93 & $+$ & $+$  & 0.2-0.7 & e.g. 1\\
SMC           & 0.07& -16.99 & $+$ &   -? &     & e.g. 2\\
Scl\ dSph     & 0.08& -10.40 & $+$ &   -  &     & 3\\
Car\ dSph     & 0.1 & -11.60 & $+$ &   -  &     & 3\\
For\ dSph     & 0.14& -12.60 & $+$ &   -  &     & 4\\
IC\ 1613      & 0.7 & -14.20 & $+$ &   -? &     & 5, 6, 7\\
IC\ 10        & 0.8 & -15.20 & $+$ &   -? &     & 8, 9\\
NGC\ 6822     & 0.5 & -14.70 & $+$ &$\circ$&     & 5, 10, 11\\
Leo\ A        & 0.7 & -11.30 &  -  &$\circ$&     & 11 \\
NGC\ 205      & 0.8 & -16.00 &  -  &$\circ$&     & 11 \\
NGC\ 221      & 0.8 & -15.80 & $+$ &$\circ$&     & 12, 11 \\
NGC\ 147      & 0.7 & -14.80 &  -? &   -  &     & 8, 13\\
NGC\ 185      & 0.6 & -14.70 &  -? &   -  &     & 8, 13\\
NGC\ 3109     & 1.2 & -15.20 & $+$ &$\circ$&     & 14\\
NGC\ 1569     & 1.4 & -15.97 & $+$ & $+$  & 0.8 & 15, 16\\
Scl\ DIG      & 1.5 & -10.42 & $+$ &$\circ$&    & 17\\
NGC\ 625      & 1.5 & -14.69 & $+$?& $+$  & 0.2 & 18\\
Ho\ II        & 3.3 & -16.67 & $+$ & -  &     & 19, 11\\
Ho\ IX        & 3.3 & -13.52 & $+$ & -  &     & 20, 11\\
NGC\ 3077     & 3.3 & -17.27 & $+$ & ?  &     & 5\\
IC\ 2574      & 3.3 & -17.26 & $+$ & $+$? & 0.5 & 5, 21, 22\\
NGC 2366      & 3.3 & -16.63 & $+$ & -  &     & 16, 9\\
KDG\ 061      & 3.3 & -12.99 & $+$?& -  &     & 13\\
UGC\ 6541     & 3.5 & -13.42 & $+$ & -? &     & 23\\
NGC\ 4449     & 3.5 & -17.82 & $+$ & $+$ & 0.2, 0.8 & 5, 24, 25, 11\\
NGC\ 4214     & 3.5 & -17.56 & $+$ & -? &     & 9\\
NGC\ 4190     & 3.5 & -13.85 & $+$ &$\circ$&     & 5, 11\\ 
NGC\ 3738     & 3.5 & -15.68 & $+$?&   -  &     & 13\\
NGC\ 5253     & 4.1 & -17.70 & $+$ & $+$  & 0.3 & 5, 26, 23, 27\\
NGC\ 5408     & 4.1 & -16.37 & $+$?&   -? & 0.5 & 28, 21, 16, 11\\
UGC\ 6456     & 4.5 & -13.49 & $+$ & $+$? &     & 29, 13\\
He\ 2-10      & 5.7 & -17.76 & $+$ &   -? & 0.5 & 21, 23, 30, 11, 31\\
NGC\ 1705     & 6.1 & -16.34 &  -  & $+$  & 0.2 & 32, 11\\
I\ Zw\ 18     & 7.9 & -14.17 & $+$ & $+$  &     & 33, 34, 21, 16\\
NGC\ 4861     & 8.4 & -16.62 & $+$ & $+$? & 0.6? & 5, 21, 23\\
NGC 1427A     & 15.2& -17.52 & $+$?& $+$? & 0.6? & 35\\
\hline
\hline
\end {tabular}
\end{center}
\end{table}

\begin{table}
{\small\it
(1) Snowden \& Petre (1994), (2) Snowden (1999), (3) Zinnecker et al. (1994), 
(4) Gizis et al.\ (1993), (5) Fabbiano et al. (1992), (6) Eskridge (1995), 
(7) Lozinskaya et al.\ (1998), (8) Brandt et al. (1997), (9) Roberts \&
Warwick (2000), (10) Eskridge \& White (1997), (11) Colbert \& 
Mushotzky (1999), (12) Eskridge et al.\ (1996), (13) Lira et al. (2000), 
(14) Kahabka et al.\ (2000), (15) Heckman et al.\ (1995), 
(16) Stevens \& Strickland (1998b), (17) Burstein et al.\ (1997), 
(18) Bomans \& Grant (1998), (19) Zezas et al.\ (1999), (20) Miller (1995), 
(21) Fourniol et al.\ (1996), (22) Walter et al. (1998), 
(23) Stevens \& Strickland (1998a), (24) Bomans et al.\ (1998), 
(25) Vogler \& Pietsch (1998), (26) Martin \& Kennicutt (1996),
(27) Strickland et al.\ (1999), (28) Fabian \& Ward (1993), 
(29) Papaderos et al.\ (1994), (30) Dickow et al. (1996), 
(31) M{\'e}ndez et al.\ (1999), (32) Hensler et al.\ (1998), 
(33) Martin (1996), (34) Bomans (1999), (35) Hilker et al.\ (1997)
}
\\
{\small\it
The distances and luminosities of the Local Group galaxies are taken from  
the compilation of Mateo (1998). For the other galaxies the data are 
taken from Schmidt et al. (1993) and the LEDA database, but all distances and 
M$_B$ were recomputed relative to the HST based distance modulus of 
the Virgo cluster (Ferrarese et al. 1996).   In column 3 and 4 a "$+$" denotes 
detection, "-" denotes non-detection, and "$\circ$" where no analysis of the 
data concerning this type of emission was performed or reported.  An 
additional "?" denotes uncertain or contradicting evidence. Column 5 gives the 
temperature of the probable diffuse gas, when determined.
}
\end{table}

An additional problem to interprete the X-ray emission of dwarf galaxies 
is the fact that both the EINSTEIN and the ROSAT samples are preselected
to be galaxies which are for one or another reason interesting to the 
principal investigators of the original pointed observations or by 
chance inside the field of view.  Therefore the samples are far from 
statistically well behaved.  One attempt to avoid this problem has been 
undertaken by Schmidt et al. (1996).  They used the ROSAT All Sky Survey 
(RASS) data and their catalog of nearby galaxies (radial velocity 
$< 500$ \kms) as input sample.  Unfortunately the exposure time of the RASS 
is only a few 100 sec on average, yielding a quite low sensitivity.  

The next step in X-ray observations of dwarf galaxies came with the ASCA 
satellite (Tanaka et al.\ 1994) which first employed CCD detectors in for 
X-ray astronomy, 
leading to a much higher spectral resolution (compared to the imaging 
proportional counters in EINSTEIN and ROSAT) combined with good sensitivity.
Drawback of ASCA is its low spatial resolution of about 1.5\arcmin, compared 
to about 25\arcsec\ for ROSAT PSPC or 5\arcsec\ for ROSAT HRI (detector 
without energy resolution). 
This low spatial resolution limits ASCA data effectively to be only an 
integrated spectrum of a dwarf galaxy without the possibility to distinguish 
between point sources and extended diffuse emission.  The other problem of 
the ASCA data is 
the relatively hard band-path of 0.5 to 10 keV, versus 0.1 to 2.4 keV of
ROSAT. This forces the combined analysis of ROSAT and ASCA datasets for 
analysis of relatively low temperature plasma.  
These limitations lead to only 2 dwarf galaxies analyzed with ASCA 
up to now, NGC\ 1569 (Della Ceca et al.\ 1996) and NGC\ 4449 (Della Ceca et
al.\ 1997).  

All objects emitting relatively soft (low energy) X-rays share an 
observational problem.  The low energies are strongly absorbed by the 
galactic N$_H$, which implies, that studies of the soft diffuse emission 
is only possible in dwarf galaxies with sufficiently low foreground N$_H$.
The imposes an additional strong and unavoidable selection bias on the X-ray
properties of dwarf galaxies.  
For example, NGC\ 1569 does not show a very soft X-ray emission, while 
NGC\ 4449 does. The difference may be intrinsic to the galaxies, but since 
the foreground N$_H$ of NGC\ 1569 is much higher, it is equally possible that 
all soft X-ray emission of NGC\ 1569 is just absorbed.  

\begin{figure}[t]
\epsfxsize=0.80\textwidth
\centerline{\epsffile{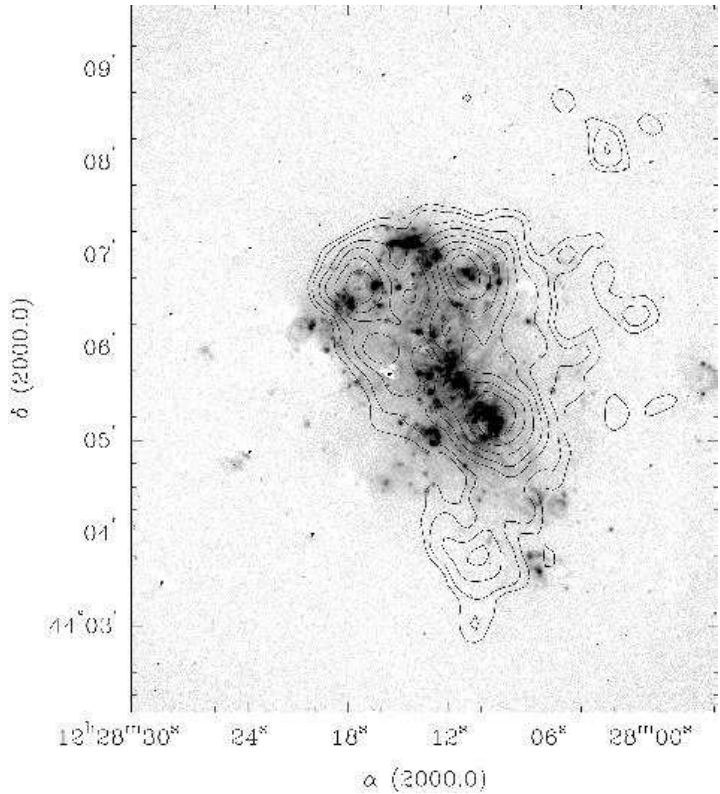}}
\caption{Pure \Ha\ image of the irregular galaxy NGC\ 4449 with contours of 
the ROSAT soft and medium band X-ray emission overplotted.} 
\label{fig:ngc4449_x}
\end{figure}

\subsection{Diffuse gas and outflows}

As we saw in section 2, the natural results of some locally increased 
star formation rate is a superbubble, which expands in the neutral 
interstellar medium of the host galaxy.  The evolution of the superbubble 
inside a dwarf galaxy is different since the conditions are different from 
that in a spiral galaxy.  Due to the small size of the dwarf galaxy, 
a big association takes up a significant part of the host galaxy. Additionally
dwarf galaxies tend to have larger OB associations than spiral galaxies when 
scaled to the galaxy sizes (Elmegreen et al.\ 1994).  The 
bubble, which has to develop will not be sheared in the solid-body 
rotation field of the dwarf galaxy and expands undisturbed to larger size due
to the thicker HI layer (e.g. Skillman 1995).  As soon the shell starts to 
break out, the shallow gravitational potential well of the dwarf galaxy makes 
it more likely for the bubble to reach large distances from the host galaxy.  
The low metallicity inhibits cooling which helps maintain a high pressure in 
the bubble (before breakout).  

The catch in this scenario is the extended dark matter halo of dwarf 
galaxies.  If the maximum disk interpretation of 
rotation curves of dwarf galaxies is correct (Swaters 1999), 
dwarf galaxies are even more dark matter dominated than disk galaxies. 
This implies, that we need some assumptions on shape and total size 
of the dark matter halo to estimate if vented out gas will stay in the 
gravitational potential or will be lost into the intergalactic medium. 
Due to the large dark matter fraction of dwarf galaxies this correction 
is much more critical than in disk galaxies. 

Even in the small current 
sample (Tab.\,\ref{tab:x_dwarfs}) at least 6 larger and 
smaller dwarf galaxies show exactly what this picture implies:  
strong star formation and large \Ha\ shells and filaments sticking out into 
the lower halo.  Good examples are NGC\ 4449 (Bomans et al.\ 1997; Vogler \& 
Pietsch 1997) and NGC\ 1569 (Heckman et al. 1995).  Indeed, diffuse 
X-ray emission can be found inside some of the shells as demonstrated 
for NGC\ 4449 in Fig.\,\ref{fig:ngc4449_x}, but the 
integration times used for the PSPC observations turned out to be 
to small for for deriving useful spectra for determining the 
plasma conditions. Especially, the data are not of sufficient quality to 
allow the critical test if the hot gas is metal enriched compared to the 
global interstellar medium in these galaxies.

The most favorable object for the metallicity test is I\ Zw\ 18, the 
most metal poor galaxy known (e.g. Skillman \& Kennicutt 1993).  Because of 
the low metallicity of about 1/50 solar, any enrichment of the hot gas 
would make a big difference in the plasma emissivity.  Unfortunately, 
the ROSAT spectrum (Martin 1996) is again not good enough to determine the 
metallicity of the hot gas to reasonable degree of certainty (Bomans 1999). 

\begin{figure}[t]
\epsfxsize=0.80\textwidth
\centerline{\epsffile{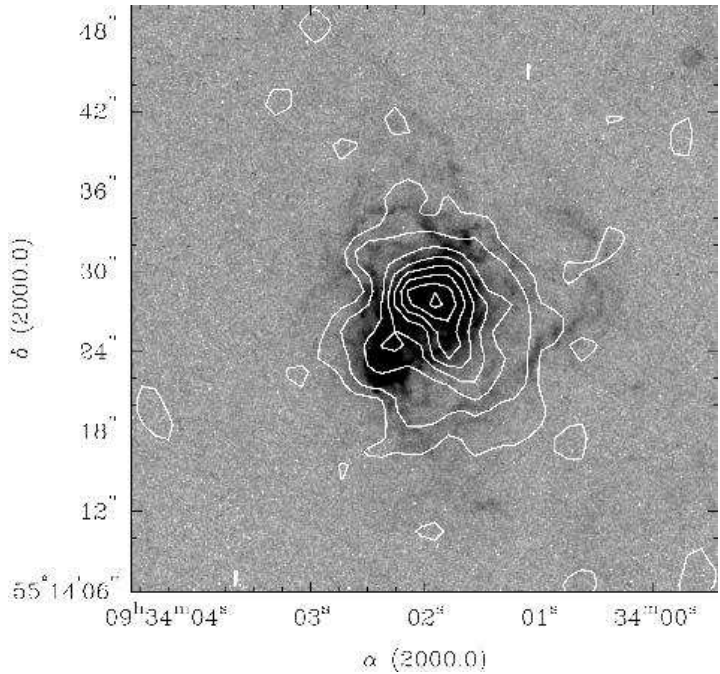}}
\caption{Continuum subtracted HST WFPC2 \Ha\ image of the dwarf galaxy 
I\ Zw\ 18 with contours of X-ray emission from ROSAT HRI overplotted.} 
\label{fig:izw18_x}
\end{figure}

Two other problems are apparent when analyzing the diffuse X-ray emission of 
dwarf galaxies:  Due to their small intrinsic size, even at moderate 
distances the galaxies represent only a few resolution elements of 
the ROSAT PSPC or even worse the ASCA SIS.  This makes contamination 
from points sources a real concern (e.g. Vogler \& Pietsch 1997).  The 
ROSAT HRI with its about 5 times better spatial resolution (but without 
spectral resolution) had a much higher background, making it 
a much inferior instrument for the detection of the low surface brightness 
diffuse emission.   Still, with long exposures, the point source content 
of the PSPC images can be checked and the brighter diffuse emission be 
studied (Bomans 1998, Strickland et al.\ 1999).  Fig.\,\ref{fig:izw18_x} 
shows a deep ROSAT HRI image of I\ Zw\ 18 as contours overlayed over an 
HST \Ha\ image. 

The other problem is the plasma itself.  If it expands rapidly into the 
lower density halo, the gas cools adiabatically, but roughly maintains 
its ionization patter, recombination lacks behind cooling (Breitschwerdt \& 
Schmutzler 1999).
Therefore the standard analysis of X-ray spectrum using the coronal 
equilibrium plasma codes (Raymond \& Smith 1977; Mewe et al.\ 1995) 
does not necessarily determine the physical parameters correctly, even 
if one has a very good spectrum.  Much higher spectral resolution together 
with good signal to noise ratio is needed in the spectra to check for the 
presence and size of the effect and finally account for non-equilibrium 
conditions in the derived plasma properties. While the spectral 
resolution of ASCA helps, it could not produce sufficiently high 
quality spectra of the diffuse gas (della Ceca et al. 1997, 1998).

\subsection{Diffuse hot halos}

With outflows of hot gas out of dwarf galaxies observed in at least 6 
dwarf galaxies with strong star formation (Tab.\,\ref{tab:x_dwarfs}), 
the question arises what the ultimate fate of this gas will 
be.  This 
problem is firmly linked to the understanding of chemical evolution of 
dwarf galaxies due to the metal-enriched nature of the hot gas.
Current theories of chemical evolution of dwarf galaxies postulate the 
presence of galactic winds to account for both the star formation history 
and the current (low) metallicities of dwarf galaxies (e.g. Matteuchi \&
Chiosi 1983; Hensler \& Burkert 1990).  
Also the best explanation for the 
very homogeneous metallicity level of practically all dwarf galaxies 
(Kobulnicky \& Skillman 1996; 1997) seems to be 
outflows of hot metal enriched gas, which stays in the halo for a while 
and then drizzles down to the galaxy again (e.g. Pantelaki \& Clayton 1987; 
Roy \& Kunth 1995; Tenorio-Tagle 1996).  For more arguments in favor and 
against this picture see Skillman (1997).

A hot gaseous halo around a dwarf galaxy should be 
expected, either 
from a galactic wind (where most of the mass really leaves the potential 
well of the dwarf galaxy into the intergalactic space) or from outflows 
(which pump material in the halo but with velocities below escape 
velocity of the potential well of the dwarf galaxy).  How this halo would look 
like is highly depending on the properties of the gravitational potential 
and the history of star formation both in the temporal 
and spatial domains.  It is also depending on the metallicity, density, and 
topology of the gas itself due to the cooling processes.  In any case, 
hot, low metallicity, low density gas cools slowly (e.g. Boehringer \& Hensler 
1989). One more specific hint can be extracted from the simulations 
of Mac Low \& Ferrara (1999) which showed that in a relatively realistic 
gravitational potential the hot gas (and therefore the metals) can reach 
large distances from the host galaxy.  They also showed that the structure 
may be more like pockets of hot gas and less a smooth hot halo.
This last conclusion should be less dependable, since the cooling of the 
hot gas is critical here and had to be treated in a relatively simplistic 
manner in these simulations.  No other set of simulations is available in 
the literature yet, which follows the evolution 
of bubbles and outflows long enough to study the properties and the ultimate 
fate of the hot gas in dwarf galaxies.
Still, one can use the results on hot halos of low velocity dispersion 
elliptical galaxies to make rough predictions. Adiabatic compression would 
keep the halo gas hot and the calculations predict 
temperatures in the order of 0.2 keV for galaxies with masses similar 
to the LMC and below (e.g. Nulsen et al. 1984).

On the observational side, two attempts have been made to look for 
such extended hot halos around dwarf galaxies. The first 
program used pointed PSPC observations  of 3 irregular galaxies 
with distances 
in the order of 10 Mpc (Bothun et al. 1994).  In the two exposures which 
reach decent sensitivity the target galaxies where detected, but only 
as sources with hard spectrum and without spatial extension. 
The sensitivity 
of the observations results in a non-detection of soft ($\sim 0.2$ keV)
extended halo with masses of $10^9$ \Msun\ or above.
The other search for hot extended halos around dwarf galaxies uses the 
ROSAT archive to select all star forming dwarf galaxies within a distance 
of 6 Mpc which had PSPC observations (pointed and serendipitous) and low 
foreground N$_H$ column density.  The sample 
contained 49 galaxies with PSPC data, but no large hot halo was detected 
in the first pass of data analysis (Bomans 1998). The 
detectable mass of hot gas in a spherical hot halo with 10 kpc radius 
around a galaxy at 6 Mpc distance is about 10$^6$ \Msun\ in a typical 
10 ksec exposure. The survey detected diffuse hot gas in several dwarf 
galaxies, but it is always located quite close to the galaxies (in order 
of 2 kpc), and is linked to large \Ha\ shell structures.

If the basic assumptions are right, then three conclusions can 
be drawn: 1) The hot gas expands to very large distances (or is lost to the 
intergalactic space) giving it  such low surface brightness that the 
present ROSAT PSPC data are not sensitive enough to detect it. 
2) The gas is highly clumped as implied by Mac Low \& Ferrara (1999) and not a 
smooth halo. Such hot gas pockets would be missed by the analysis methods 
tailored to detect extended diffuse emission. 3) The gas could be at 
such low temperature that the peak of the spectrum is located outside the 
ROSAT energy range.  The stringent limits for the EUV 
flux of large star forming galaxies using the ROSAT WFC (Read \& Ponman 1995) 
make at least the third possibility quite unlikely.

An alternative way to search for hot plasma in the halos 
of nearby dwarf galaxies was tested by Bowen et al. (1997).  They used 
HST high-dispersion UV spectra to 3 QSO in the background of the 
dwarf spheroidal galaxy Leo I.  No highly ionized gas was detected, again 
hinting that the possibly missing hot halos are not at temperatures 
of a few $10^5$ K, as traced by the CIV UV lines.

\subsection{Integral spectra}
Up to now, only for two dwarf galaxies X-ray spectra of higher spectral 
resolution (the CCD detectors of ASCA): NGC\ 1569 (Della Ceca et al. 1996) 
and NGC\ 4449 (Della Ceca et al. 1997) have been analyzed.   
In both cases the signal to 
noise ratio is not good enough to measure reliable metallicities of the 
gas using lines/line complexes.  Still, the much larger spectral range 
allowed a look at the higher energy part of the integrated 
X-ray spectrum of the two dwarf galaxies.  
The spectra showed in both cases the need for at least 2 components, a soft 
one with kT $\sim 0.8$ keV and a hard component 
with kT $\sim 3.5$ keV.  In the case of NGC\ 4449 an additional very soft 
component (kT $\sim 0.2$ keV) is needed, too, consistent with the ROSAT-only 
analyzes of Bomans et al.\ (1997) and Vogler \& Pietsch (1997).
The hard component is best interpreted as a mix of young supernova remnants 
and X-ray binaries, while the soft and especially the very soft components  
is largely due to diffuse hot gas, consistent with the extended nature 
of the emission on the ROSAT images.  It is worth to note, that the 
ASCA spectra of both observed galaxies could only be detected out to 
$\sim 6$ keV, making such dwarf galaxies only weak contributors to the 
hard X-ray background (della Ceca et al. 1996).

\section{Diffuse warm gas}
Structure, occurrence, and ionization of the diffuse warm gas and especially 
the presence of large ionized shells in dwarf galaxies were discussed in a 
number of recent publications (e.g. Martin 1997, 1998; Hunter \& Gallagher 
1997; Hunter et al.\ 1993).  Here I will concentrate mostly on the dynamics 
of the ionized gas and its implications for the fate for the hot gas.


First it is important to note here, that the warm ionized gas cannot 
be used for the metallicity determination of the outflows.  We have seen 
that there are signs of at least some mixing at the boundary layers between 
hot and cold gas and therefore a metal enrichment of warm gas.  This makes the 
observation of the optical emissions 
lines of the shells and outflows tempting as alternative way to estimate 
the metal content of the outflows.  Still the spectra cannot 
be treated with the usual methods as used for normal HII region.  The line 
ratios clearly indicate a complex interplay of normal stellar photoionization, 
photoionization by a diffuse photon field, shock ionization, 
photoionization by X-ray, and even turbulent mixing layers (Hunter \& 
Gallagher 1997; Martin 1998, T\"ullmann \& Dettmar 2000).  With insufficient 
information to even determine the ionization mechanism(s), a direct 
derivation of ionic abundances from optical emission lines for the diffuse 
ionized gas  is out of the question (at least for now).  
A helpful alternative method may be the use of interstellar UV absorption 
lines to background QSOs.  Unfortunately, no usable chance alignment of a 
dwarf galaxy outflow with a sufficiently bright background source has been 
found.
 
When one finds a dwarf galaxy showing large \Ha\ shells, which extend to 
a size larger than the stellar body of the dwarf galaxy, the first 
critical question is if this ionized gas is really expanding away from 
the dwarf galaxy.  Observationally the dynamics of the ionized 
gas can be studied either with high-dispersion, long-slit spectroscopy, or 
using a Fabry-Perot interferometer.  While the long-slit approach 
only uses one spatial axis, it is stable against changes in instrumental and 
weather conditions,
relatively easy to reduce, and can provide high spatial and spectral 
resolution ($\sim 10$ \kms). The one-dimensional spatial axis allows also a 
very good 
sky subtraction and therefore sensitivity to faint features. Producing a
2-d map on the other hand, requires offsetting the slit and is therefore 
very telescope time intensive. The scanning 
Fabry-Perot approach delivers a real data cube with two spatial and a 
spectral axis, but is somewhat susceptive to changes in the observing 
conditions and the data relatively hard to handle. Unfortunately, no 
Fabry-Perot dataset on dwarf galaxies has delivered yet at the same time 
the high spectral resolution and sensitivity of the best long-slit data. 
The alternative use of the Fabry-Perot as spectrometer
delivers very high sensitivity at the cost of low spatial resolution and is 
therefore of only limited use for the study of outflows. 

\begin{figure}[t]
\epsfxsize=0.80\textwidth
\centerline{\epsffile{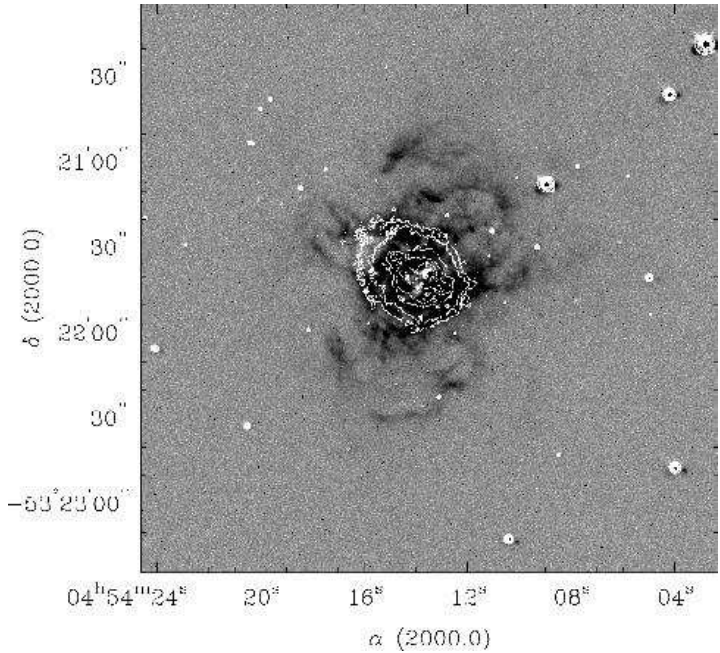}}
\caption{Continuum corrected VLT \Ha\ image of NGC\ 1705 with contours of the 
continuum emission overplotted.} 
\label{fig:ngc1705_h}
\end{figure}

As an example I present here an analysis of the galaxy NGC\ 1705. 
NGC\ 1705 contains several large \Ha\ shells (Meurer et al.\ 1992), two of 
which show up in a deep ROSAT PSPC pointing as very soft, diffuse X-ray 
sources (Hensler et al.\ 1998).    
The data presented here were taken with the ESO VLT and the ESO NTT and are 
going to be discussed in detail in Bomans et al. (2001b, in prep.).
In Fig.\,\ref{fig:ngc1705_h} the continuum subtracted \Ha\ image of 
NGC\ 1705 is shown with contours of the Gunn-r continuum image overlayed. 
The body of the galaxy is relatively smooth, indicating a long phase of 
low star formation activity, and one bright knot, which harbors 
a young globular cluster (Melnick et al.\ 1985;  Ho \&
Filippenko 1996).
The about 1.8 kpc large stellar body (Holmberg diameter)  
of the galaxy is nearly filled with HII regions and large ionized shells 
and filaments extend radially out to more than 2.5 kpc from the center of the 
dwarf galaxy.  The structure appears roughly bipolar, consistent with 
hydrodynamical simulations, but shows a very complex substructure with 
interlocking and overlapping individual shells and filaments. One should keep 
in mind that we see in the image a 2-d projection of the 3-d structure, 
which may explain at least some of the shells-inside-shells as 
line-of-sight projection. 
The currently available HI synthesis map is of to low spatial resolution to 
study the detailed dynamic of the gas, but shows that NGC\ 1705 is embedded 
in a large HI envelope.

\begin{figure}[t]
\epsfxsize=0.5\textwidth
\centerline{\epsffile{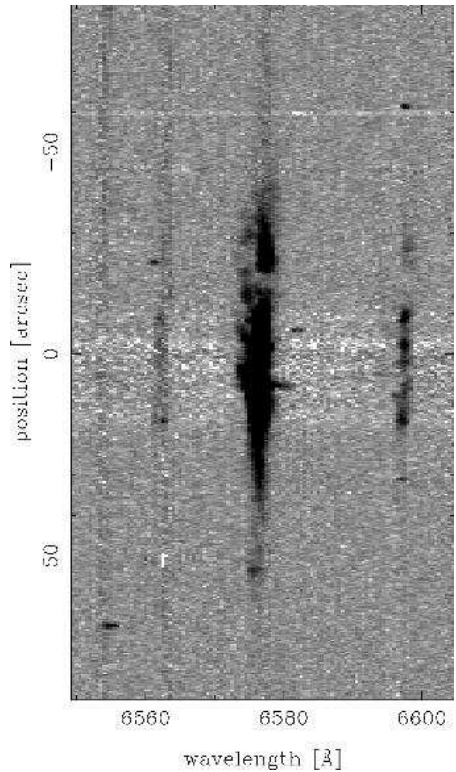}}
\caption{High-dispersion spectrogram of one slit position through NGC\ 1705.} 
\label{fig:ngc1705_spec}
\end{figure}

Fig.\,\ref{fig:ngc1705_spec} shows a long-slit spectrum taken with a position 
angle of $310\deg$, 
roughly aligned to the central star cluster of NGC\ 1705 and the bright 
foreground star in the north-west.  The spectrogram runs top to bottom 
from north-west to south-east and shows a spectral range of 60\AA\ centered 
on the \Ha\ line and the two [NII] lines.  Note that the lines are 
redshifted due to the radial velocity of NGC\ 1705 and the [NII] lines 
quite weak due to the sub-solar metallicity of NGC\ 1705.  
Just below the bright star (the residuum from continuum subtraction running 
as white line left to right near the top the the spectrogram) the \Ha\ line 
splits into two components and merges again at the beginning of the stellar 
body. This Doppler ellipse is the 
sign of an expanding bubble and coincides exactly with the bright loop 
in the north-west. The expansion velocity is 75 \kms.  The \Ha\ line 
maintains a complex structure over the whole stellar body of NGC\ 1705, not 
surprisingly 
given the complicated structure of the \Ha\ emission in this region visible 
in Fig.\,\ref{fig:ngc1705_h}.  At the bottom of the spectrogram the \Ha\ line
does not show a line split, but the width of the line is large, implying 
motions with 
velocities below the spectral resolution of $\sim 30$ \kms.  Interestingly, 
the slit cuts through another large shell, which is, contrary 
to the shells in the north-west, not detected in the deep ROSAT image.  
It is tempting to make a link between the lower expansion velocity of 
the shell in the south-east and the missing X-ray emission.  
Similar velocities are seen also in an spectrum with a position angle of 
$230\deg$ (Marlowe et al.\ 1995).

Still, the situation is a bit more complicated, since neither the 75 \kms\ 
nor the 30 \kms\ shock speed would give rise to X-ray emission.   
It requires about 200 \kms\ to get post-shock temperatures above $10^6$ K 
(McKee 1987). Simple projection effects are not likely to make a large effect 
for expanding bubbles.  The probable explanation comes from the 
inherent selection effect of using the warm ionized gas as tracer of 
the flow: the gas detected with the spectrometer is the gas with highest 
surface brightness and therefore density, which is the least probable to 
show the highest velocities.  We apparently measure the large scale expansion 
of the superbubble driven by the overpressure of the hot gas inside, 
which is heated by the supernovae.  As we have seen in the case of 
N51D, a significant part of the X-ray emission results from recent supernovae 
shocks hitting the shell walls.  Therefore there is a link between the 
temperature of the hot gas and the expansion velocity, but it is not 
the simple post-shock temperature from the large-scale shock, hitting the 
surrounding interstellar medium, but local heating near the inner shell wall. 
Indeed, we detect hot gas inside the shell walls not outside 
(Bomans et al. 2001b, in prep.).  

This is also consistent with the missing correlation of the X-ray temperature 
with global galaxy parameters (Tab.\,\ref{tab:x_dwarfs}). The visible 
clustering of the X-ray temperature of the diffuse gas near 0.2 and 0.8 keV 
could be an effect of the location of the local minima visible in the cooling 
curve (Boehringer \& Hensler 1989) at low metallicity.

The shells of NGC\ 1705 are clearly expanding out of the galaxy and they are 
(at least in part) filled with hot gas.  The question is now, what will be the 
fate of this bubbles?  The faintest shells and filaments visible 
on our VLT \Ha\ image reach the rim of the HI envelope of NGC\ 1705.
If they are still expanding at this radius, they are likely going to speed 
up when hitting the gas density gradient.  Unfortunately, the filaments are 
very faint and it is hard to measure their radial velocity.  
Even with a measured velocity, the lack of HI also means that the rotation 
curve and therefore the potential of the galaxy is not well constrained there.
These problems are generic to the analysis method and should be kept in 
mind when evaluating similar analyzes of the mass loss from dwarf galaxies 
(e.g. Martin 1998).

There is another effect, which may be important for the processes discussed 
here. Ordered magnetic fields, if present, could have a large influence on the 
conditions of the outflows and the kinematics of the warm and hot gas. 
Unfortunately, the knowledge about magnetic 
fields in dwarf galaxies is very sparce.  From theoretical considerations 
no large-scale ordered magnetic fields are to be expected in dwarf galaxies, 
but at least the LMC (Klein et al. 1993) and NGC\ 4449 (Chy{\.z}y et al. 2000) 
do show such magnetic fields.  Still, both galaxies are at the upper mass 
and luminosity boundary for dwarf galaxies and the conditions in lower mass  
dwarf galaxies are unexplored yet.

\section{Conclusions and Outlook}

The link between superbubbles in the interstellar medium and diffuse X-ray 
emission has been clearly demonstrated (e.g. Chu \& Mac Low 1990) and even 
some larger bubbles have been shown to be filled with hot gas (e.g. Wang 
et al. 1991, Bomans et al. 1994).  The detailed physics of such 
bubbles is still ill observed.  We do not know the metallicity of the hot gas, 
the ionization condition (equilibrium or non-equilibrium), the details 
of the interaction between hot gas and cool shell, or the presence and 
properties of cold cloudlets inside 
the hot cavity.   In dwarf galaxies, the presence of extended hot gas 
could be proven and even that parts of this gas is located far away from 
its possible sites of creation (e.g. Heckman et al. 1995, Bomans et al. 1997).
The number of such studies is still very small (see Tab.\,\ref{tab:x_dwarfs})
and limited to galaxies with recent strong star formation.  No dwarf galaxy 
with diffuse hot halo is known up to now. Likewise no dwarf galaxy with 
extended X-ray emission but currently low star formation rate has been found 
up to now.  
Clearly we do not know from observation, what the fate of the hot gas 
will be.  Using the kinematics of the warm ionized gas together 
with the HI rotation curve one can estimate if hot gas inside 
a bubble will escape the potential well of a dwarf galaxy (e.g. Martin 1998), 
but the uncertainties are large, partly due to the big uncertainties 
in determining the dark matter potential (e.g. Swaters 1999), partly 
due to the current inability to measure the velocity of the faintest
shells and the warm ionized gas with highest velocities (see Section 4).

The determination of the physical parameters of the hot gas are hampered by 
the quality of the X-ray spectra and the contamination with point sources. 
The new X-ray telescopes (XMM-NEWTON and 
CHANDRA) are now in orbit and are working well.  
Especially XMM-NEWTON promises large improvements in the quality of the 
spectra due to its unmatched sensitivity and good spatial resolution of 
$\sim 15$\arcsec. CHANDRA is especially well suited for the study of the 
point source population due to its very good spatial resolution of 
$\sim 0.5$\arcsec, but lower sensitivity. Unfortunately, 
most of the diffuse X-ray emission is expected to be in the very soft 
X-ray regime (see Tab.\,\ref{tab:x_dwarfs}), where both satellites are 
hard to calibrate.  Still, at least for the dwarf galaxies with brighter 
diffuse X-rays the new instruments should allow to analyze the plasma 
conditions and measure metallicity of the hot gas.  This will provide a big 
step forward in testing the current dwarf galaxy evolution scenarios.

For the diffuse warm gas, the near futures looks also promising, with 
several 8-10m class telescopes currently coming online.  This will enable 
us to study the kinematics of the outflows even at low surface brightness, 
hunt for very high velocity gas and presumably extremely faint outer halo 
gas using emission lines and quasar absorption lines.   
These methods should also help to better determine the mass distribution and 
total mass of dwarf galaxies, leading to improved estimates on the escape 
fraction of the hot gas and therefore the metals.

While we started to answer the question if diffuse warm and especially hot gas 
is a common constituent of the interstellar matter in dwarf galaxies, 
the detailed physics of its creation, and evolution, as well as the links to 
global evolution of dwarf galaxies and the intergalactic medium 
have only slightly been touched yet.  There should be exiting years to 
come!

\subsection*{Acknowledgments}
The author is very grateful to K. Weis for critically and thoroughly 
reading the manuscript and many stimulating discussions. Furthermore, 
the author likes to thank Y.-H. Chu, M. Dahlem, K. Dennerl, 
R.-J. Dettmar, W.J. Duschl, G. Hensler, U. Hopp, N. Junkes, M.-M. Mac Low, 
P. Papaderos, G. Richter, E.D. Skillman, R. T\"ullmann,  M. Vogler, and 
M. Urbanik for discussions on many facets of dwarf galaxies and outflows. 
This work was supported by Verbundforschung grant 50 OR 99064 and 
the Bonn-Bochum Graduiertenkolleg "Magellanic Clouds and other Dwarf Galaxies".
This research has made use of the NASA/IPAC Extragalactic Database (NED) 
which is operated by the Jet Propulsion Laboratory, California Institute of 
Technology, under contract with the National Aeronautics and Space 
Administration. This work has also made use of the NASA's Astrophysics Data 
System Abstract Service and the LEDA database (http://leda.univ-lyon1.fr).

\vspace{0.7cm}
\noindent
{\large{\bf References}}

{\small
\bref
Boehringer H., Hensler G. 1989, A\&A 215, 147

\bref
de Boer K.S., Nash A.G. 1982, ApJ 255, 447 (erratum ApJ 261, 747)

\bref
Bomans D.J. 1998, in: Galactic Halos, ed. D. Zaritsky, ASP Conf. Ser. 136, 138

\bref
Bomans D.J. 1999, in: Chemical Evolution from Zero to High Redshift, eds.: 
J. Walsh, M. Rosa, ESO Astrophysics Symposia, 162

\bref
Bomans D.J. 2000, in: `Evolution of Galaxies. I - Observational Clues',
ed. G. Stasinska, Ap\&SS, in press

\bref
Bomans D.J., Dennerl K., K\"urster M. 1994, A\&A 283, L21

\bref
Bomans D.J., Chu Y.-H., Hopp U. 1997, AJ 113, 1678

\bref
Bomans D.J., Grant M.-B. 1998, AN 319, 26

\bref
Bothun G.D., Eriksen J., Schombert J.M., 1994, AJ 108, 913

\bref
Bowen D.V., Tolstoy E., Ferrara A., Blades J.C., Brinks E. 1997, ApJ 478, 530

\bref
Brandt W.N., Ward M.J., Fabian A.C., Hodge P.W. 1997, MNRAS 291, 709

\bref
Breitschwerdt D., Schmutzler T. 1999, A\&A 347, 650

\bref
Burstein D., Jones C., Forman W., Marston A.P., Marzke R.O. 1997, ApJS 111, 163

\bref
Castor J., Weaver R., McCray R. 1975, ApJ 200, L107

\bref
della Ceca R., Griffiths R.E., Heckman T.M., MacKenty J.W. 1996, ApJ 469, 662

\bref
della Ceca R., Griffiths R.E., Heckman T.M. 1997, ApJ 485, 581

\bref
Chu Y.-H. 1995, RexMexAAp Conf. Ser. 3, 153

\bref
Chu Y.-H., Mac Low M.-M. 1990, ApJ 365, 510

\bref
Chu Y.-H., Mac Low M.-M., Garcia-Segura G., Wakker B., Kennicutt R.C. 1993, 
ApJ 414, 213

\bref
Chy{\.z}y K.T., Beck R., Kohle S., Klein U., Urbanik M. 2000, A\&A 355, 128

\bref
Colbert E.J.M., Mushotzky R.F. 1999, ApJ 519, 89

\bref
Dickow R., Hensler G., Junkes N. 1996, in: The Interplay
Between Massive Star Formation, the ISM and Galaxy Evolution, eds.: 
D. Kunth, B. Guiderdoni, M. Heydari-Malayeri, T.X. Thuan, Editions
Frontiers, 583

\bref
Elmegreen D.M., Elmegreen B.G., Lang C., Stephens C. 1994, ApJ 425, 57

\bref
Eskridge P.B. 1995, PASP 107, 561

\bref
Eskridge P.B., White R.E., Davis D.S. 1996, ApJL 463, 59

\bref
Eskridge P.B., White R.E. 1997, AJ 114, 988

\bref
Fabbiano G., Kim D., Trinchieri G. 1992, ApJS 80, 531

\bref
Fabian A.C., Ward M.J. 1993, 263, 51

\bref
Ferrarese L., Freedman W.L., Hill R.J., et al. 
1996, ApJ 464, 568

\bref
Freyer T., Hensler G. 2000, RevMexAAp Conf. Ser. 9, 187

\bref
Fourniol N., Pakull M., Motch C. 1996, in: Roentgenstrahlung
from the Universe, eds.: H.U. Zimmermann, J.E. Tr\"umper, H. Yorke, MPE Report
263, 375

\bref
Gerola H., Seiden P.E., Schulman, L.S. 1980, ApJ 242, 517

\bref
Giacconi R., Branduardi G., Briel U., et al. 
1979, ApJ 230, 540

\bref
Gizis J.E., Mould J.R., Djorgovski S. 1993, PASP 105, 871

\bref
Haffner L.M., Reynolds R.J., Tufte S.L. 1999, ApJ 523, 223

\bref
Heckman T.M., Dahlem M., Lehnert M.D., Fabbiano G., Gilmore D., Waller W.H. 
1995, ApJ 448, 98

\bref
Hensler G., Burkert A. 1990, ApSS 171, 231

\bref
Hensler G., Dickow R., Junkes N., Gallagher J.S. 1998, ApJL 502, 17

\bref
Hensler G., Rieschick A. 1999, in: Chemical Evolution from Zero to
High Redshift, eds. J. Walsh, M. Rosa, ESO Astrophysics Symposia, 166

\bref
Hilker M., Bomans D.J., Infante L., Kissler-Patig M. 1997, A\&A 327, 562

\bref
Ho L.C., Filippenko A.V. 1996, ApJ 472, 600

\bref
Hughes J.P., Hayashi I., Koyama K. 1998, ApJ 505, 732

\bref
Hunter D.A., Hawley W.N., Gallagher J.S. 1993, AJ 106, 1797

\bref
Hunter D.A., Gallagher J.S. 1997, ApJ 475, 65

\bref
Kahabka P., Puzia T.H., Pietsch W. 2000, A\&A 361, 491

\bref
Kennicutt R.C., Bresolin F., Bomans D.J., Bothun G.D., Thompson I.B. 1995, 
AJ 109, 594

\bref
Klein U., Haynes R.F., Wielebinski R., Meinert D. 1993, A\&A 271, 402
       
\bref
Kim S., Staveley-Smith L., Dopita M.A., Freeman K.C., Sault R.J., 
Kesteven M.J., McConnell D. 1998, ApJ 503, 674

\bref
Kobulnicky H.A., Skillman E.D. 1996, ApJ 471, 211

\bref
Kobulnicky H.A., Skillman E.D. 1997, ApJ 489, 636

\bref
Lasker B.M. 1980, ApJ 239, 65

\bref
Lira P., Lawrence A., Johnson R.A. 2000, MNRAS 319, 17

\bref
Lozinskaya T.A., Silchenko O.K., Helfand D.J., Goss W.M. 1998, AJ 116, 2328

\bref
Pantelaki I., Clayton D.D. 1987, in: Starbursts and Galaxy Evolution, 
Editions Frontieres, 145

\bref
McKee C.F. 1978, in Spectroscopy of Astrophysical Plasmas, eds. A Delgarno, 
D. Layzer, Cambridge Univ. Press, 226
 
\bref
Mac Low M.-M., McCray R. 1988, ApJ 324, 776

\bref
Mac Low M.-M., McCray R., Norman M.L. 1989, ApJ 337, 141

\bref
Mac Low M.-M., Ferrara A. 1999, ApJ 513, 142

\bref
Marlowe A.T., Heckman T.M., Wyse R.F.G., Schommer R. 1995, ApJ 438, 563

\bref
Martin C.L. 1996, ApJ 465, 680

\bref
Martin C.L. 1997, ApJ 491, 561

\bref
Martin C.L. 1998, ApJ 506, 222

\bref
Martin C.L., Kennicutt R.C. 1995, ApJ 447, 171

\bref
Mateo M.L., 1998, ARA\&A 36, 435

\bref
Matteucci F., Chiosi C. 1983, A\&A 123, 121

\bref
Meaburn J. 1980, MNRAS 192, 365

\bref
Melnick J., Moles M., Terlevich R. 1985, A\&A 149, L24

\bref
M{\'e}ndez D.I., Esteban C., Filipovi{\'c  M.D., Ehle M., Haberl F., Pietsch 
W., Haynes R.F. 1999, A\&A 349, 801

\bref
Meurer G.R., Freeman K.C., Dopita M.A., Cacciari C. 1992, AJ 103, 60

\bref
Meurer G.R., Staveley-Smith L., Killeen N.E.B. 1998, MNRAS 300, 705

\bref
Mewe R., Kaastra J.S., Liedahl D.A. 1995, Legacy 6, 16 

\bref
Miller B.W. 1995, ApJL 446, 75

\bref
Nulsen P.E.J., Stewart G.C., Fabian A.C. 1984, MNRAS 208, 185

\bref
Oey M.S., Smedley S.A. 1998, AJ 116, 1263

\bref
Papaderos P., Fricke K.J., Thuan T.X., Loose H.-H. 1994, A\&A 291, L13

\bref
Raymond J.C., Smith B.W. 1977, ApJS 35, 419

\bref
Read A.M., Ponman T.J. 1995, MNRAS 276, 1327

\bref
Roberts T.P., Warwick R.S. 2000, MNRAS 315, 98

\bref
Roy J.-R., Kunth D. 1995, A\&A 294, 432

\bref
Schmidt K.-H., Priebe A., Boller T. 1993, AN 314, 371

\bref
Skillman E.D. 1997, RevMexAAp Conf. Ser. 6, 36

\bref
Skillman E.D., Kennicutt R.C. 1993, ApJ 411, 655

\bref
Skillman E.D., Bender R. 1995, RevMexAAp Conf. Ser. 3, 25

\bref
Snowden, S.L. 1999, in: New Views of the Magellanic Clouds, eds.: 
Y.-H. Chu, N. Suntzeff, J. Hesser, D.Bohlender, IAU Symp. 190, 32

\bref
Snowden S.L., Petre R. 1994, ApJL, 436, 123 

\bref
Snowden S.L., Freyberg M.J., Kuntz K.D., Sanders W.T. 2000, ApJS 128, 171

\bref
Stevens I.R., Strickland D.K. 1998a, MNRAS 294, 523

\bref
Stevens I.R., Strickland D.K. 1998b, MNRAS 301, 215

\bref
Strickland D.K., Stevens I.R. 1999, MNRAS 306, 43

\bref
Swaters R. 1999, Dark Matter in Late-Type Dwarf Galaxies, PhD Thesis, 
Univ. Groningen

\bref
Tanaka Y., Inoue H., Holt S.S. 1994, PASJ 46, L37

\bref
Tenorio-Tagle G. 1996, AJ 111, 1641

\bref
Tenorio-Tagle G., Bodenheimer P. 1988, ARA\&A 26, 145

\bref
Tomisaka K. 1998, MNRAS 298, 797

\bref
Tr\"umper J, 1993, Sci 260, 1769

\bref
T\"ullmann R., Dettmar R.-J. 2000, A\&A 362, 119

\bref
Tully R.B. 1988, Nearby Galaxies Catalog, Cambridge Univ. Press

\bref
Vogler A., Pietsch W. 1997, A\&A 319, 459

\bref
Walter F., Kerp J., Duric N., Brinks E., Klein U. 1998, ApJL 502, 143

\bref
Wang Q.D., Hamilton T., Helfand D.J., Wu X. 1991, ApJ 374, 475

\bref
Wang Q.D., Wu X. 1992, ApJS 78, 391

\bref
Weaver R., McCray R., Castor J., Shapiro P., Moore R. 1977, ApJ 218, 377 
(erratum ApJ 220, 742)

\bref
Weis K. , Duschl W.J. 1999, in: `The Magellanic Clouds and Other Dwarf 
Galaxies', eds.: Richtler T. and Braun J.M., Shaker Verlag, 303


\bref
Zezas A.L., Georgantopoulos I., Ward M.J. 1999, MNRAS 308,302

\bref
Zinnecker, H. 1994, in: Dwarf Galaxies, eds.: G.Meylan, P. Prugniel, ESO 
Conference Proceedings, 231

}
\vfill

\end{document}